\documentclass[useAMS,usenatbib]{mn2e}

\title[Rayleigh scattering cross section]
{Rayleigh Scattering Cross Section Redward of
Ly$\alpha$ by Atomic Hydrogen}
\author[Lee and Kim]{Hee-Won Lee\thanks{E-mail:
hwlee@sejong.ac.kr} and Hee Il Kim \thanks{E-mail: khi@arcsec.sejong.ac.kr
}\\
Astrophysical Research Center for the Structure and Evolution
  of the Cosmos \\
Department of Astronomy and Space Sciences, Sejong University,
Seoul, 143-747, Korea }

\begin{document}

\date{Accepted 1988 December 15. Received 1988 December 14; in original form 1988 October 11}

\pagerange{\pageref{firstpage}--\pageref{lastpage}} \pubyear{2002}

\maketitle

\label{firstpage}

\begin{abstract}
We present a low energy expansion of the Kramers-Heisenberg formula
for atomic hydrogen in terms of $(\omega/\omega_l)$, where $\omega_l$
and $\omega$ are the angular frequencies corresponding to
the Lyman limit and the incident radiation, respectively.
The leading term is proportional to $(\omega/\omega_l)^4$, which
admits a well-known classical interpretation. With higher order terms
we achieve accuracy with errors less than 4 \%
of the scattering cross sections in the region $\omega/\omega_l\le 0.6$.
In the neighboring region around Ly$\alpha$ ($\omega/\omega_l >0.6$),
we also present an explicit expansion of the
Kramers-Heisenberg formula in terms of $\Delta\omega\equiv
(\omega-\omega_{Ly\alpha})/\omega_{Ly\alpha}$. The accuracy with errors
less than 4 \% can be attained for $\omega/\omega_l \ge 0.6$
with the expansion up to the fifth order of $\Delta\omega$.
We expect that these formulae will
be usefully applied to the radiative transfer in
high neutral column density regions, including the Gunn-Peterson
absorption troughs and Rayleigh scattering in the atmospheres of giants.
\end{abstract}


\section{Introduction}

Hydrogen is the most abundant element in the universe and therefore
one may often encounter astronomical situations associated
with the radiative transfer in a region with a very high neutral
hydrogen column density $N_{HI}$. The extended atmosphere around a giant star
is such an example, where near UV photons can be significantly
scattered by atomic hydrogen (e.g. Isliker, Nussbaumer \& Vogel 1989).
Another example may be found in searches for the first objects that
are responsible for the reionization of the universe. When the
universe is still partially neutral before the completion of the
reionization process, radiation should go through 
regions with a very high neutral
hydrogen column density $N_{HI}$.

In particular, the
interactions with low energy electromagnetic waves with angular
frequency $\omega$ are simply Rayleigh
scattering, where the scattering cross section is known to be
proportional to $\omega^4$ in the limiting case where $\omega$
is much smaller than the angular frequency 
$\omega_{Ly\alpha}$ corresponding to the Ly$\alpha$ transitions.

In quantum mechanics,
the Rayleigh scattering process is described by a second-order time
dependent perturbation theory, where the scattering atom suffers level
transition twice, one associated with the annihilation of the
incident photon and the other associated with the creation of the
scattered photon. The scattering cross section is known as the
Kramers-Heisenberg formula, which is obtained by combining an infinite
sum over all the bound $np$ states with the energy $E=-E_0/n^2$ 
and an integral over all continuum
states $n'p$ with the energy $E=E_0/n'^2$, where $E_0$ is the Rydberg
energy (e.g. Sakurai 1967). Since the wave functions are analytically
known for a single electron atom, the Kramers-Heisenberg formula for
hydrogen can be written explicitly in a closed form.

However, the Kramers-Heisenberg formula is unwieldy due to the
presence of the infinitely many atomic levels contributing to the cross 
section.  In the case of a hydrogen atom in the ground state interacting with
incident radiation with $\omega$ much less than $\omega_{Ly\alpha}$, 
this inconvenience can be overcome by expanding the Kramers-Heisenberg
formula in terms of $\omega/\omega_{Ly\alpha}$.
Devoid of any resonance in the red region of Ly$\alpha$, 
the scattering cross section is a well-behaved monotonic function of $\omega$,
and the leading term is proportional to $\omega^4$, which admits an
immediate classical interpretation.
Because $\omega^4$ dependence is the limiting behaviour of
$\omega/\omega_{Ly\alpha} \ll 1$,
inclusion of higher order terms will be useful to obtain more accurate
cross section values in the red vicinity of Ly$\alpha$.
However, very near the Ly$\alpha$ resonance, the cross section is 
well approximated by a Lorentzian and hence a polynomical approximation
becomes poor. Lee (2003) introduced an expansion of the Kramers-Heisenberg
formula near Ly$\alpha$, by computing the deviation from the Lorentzian.

In this paper, we will provide and compute the accuracy of the expansions
that provide the Rayleigh scattering cross section redward of Ly$\alpha$.

\section{Calculation}
\subsection{The Kramers-Heisenberg Formula}
The interaction of electromagnetic waves with an atomic electron is
described using the Kramers-Heisenberg formula that is obtained from
the fully quantum mechanical second-order time dependent theory (e.g.
Sakurai 1969, Merzbacher 1970).
In terms of the matrix elements of the dipole operator, 
the Kramers-Heisenberg formula can be written as
\begin{eqnarray}
{d\sigma\over d\Omega}
&=& r_0^2 \left|{1\over m_e\hbar}
\sum_I \left({{\omega({\bf p}\cdot \epsilon^{(\alpha')})_{IA}
({\bf p}\cdot \epsilon^{(\alpha)})_{AI}}\over{\omega_{IA}-\omega}} 
\right.\right.
\nonumber \\ 
&-& \left.\left.  {{\omega({\bf p}\cdot \epsilon^{(\alpha)})_{IA}
({\bf p}\cdot \epsilon^{(\alpha')})_{AI}}\over{\omega_{AI}+\omega}}
\right)\right|^2r
\nonumber \\
&=& \left({r_0\omega\over m_e\hbar}\right)^2 \left|
\sum_I \left({{({\bf p}\cdot \epsilon^{(\alpha')})_{IA}
({\bf p}\cdot \epsilon^{(\alpha)})_{AI}}\over
{\omega_{IA}^2(1-\omega/\omega_{IA})}} 
\right.\right.
\nonumber \\ 
&-& \left.\left.  {{({\bf p}\cdot \epsilon^{(\alpha)})_{IA}
({\bf p}\cdot \epsilon^{(\alpha')})_{AI}}\over{\omega_{IA}^2
(1+\omega/\omega_{IA})}}
\right)\right|^2,
\nonumber \\ 
\end{eqnarray}
where $m_e$ is the electron mass,  $r_0=e^2/m_e c^2$ is 
the classical electron radius,
and $\epsilon^{\alpha}, \epsilon^{\alpha'}$ are polarization vectors 
of incident and scattered radiation respectively.
Here, $\omega_{IA}$ is the angular frequency between the intermediate state $I$
and the ground state $A=1s$. The intermediate state $I$ includes all
the bound $np$ and free $n'p$ states, and therefore the summation notation
should be interpreted as a sum over all bound $np$ states plus an integration
over all continuum $n'p$ states. 

The Wigner-Eckart theorem allows one to separate the
matrix elements of the rank 1 tensor operator $p$ into the angular part and
the radial part, where the radial part is given by the reduced matrix elements
$<f\parallel p\parallel i>$.  (e.g. Merzbacher 1970). 
Applying the Wigner-Eckart theorem, we obtain
\begin{eqnarray}
&\sum_I& {\bf p}\cdot \epsilon^{(\alpha)})_{AI}
({\bf p}\cdot \epsilon^{(\alpha')})_{AI} \nonumber \\
&=&\sum_I |<I\parallel p\parallel A>|^2
\epsilon^{(\alpha)} \cdot \epsilon^{(\alpha')}. 
\end{eqnarray}
The dot product of the polarization gives rise to the
same dipole type angular distribution as that for Thomson scattering or 
classical Rayleigh scattering (e.g. Lee \& Ahn 1998, Lee \& Lee 1997, 
Schmid 1989).

Blueward of Ly$\alpha$ the cross section shows singular behaviours at
resonances with bound excited states, which can be avoided 
introducing damping terms associated with the finite life times 
of excited states. However, redward of Ly$\alpha$, there is no resonance 
and the cross section is a well-behaved monotonic function of $\omega_i$.
We have $\omega<\omega_{IA}$ for any intermediate state $I$ redward of 
Ly$\alpha$, which allows the expansion
\begin{equation}
\left(1\pm {\omega\over\omega_{IA}}\right)^{-1}
=1\mp {\omega\over\omega_{IA}} + {\omega^2\over\omega_{IA}^2}\mp 
\cdots.
\end{equation} 

Substituting these relations into Eq.~(1), we have
\begin{eqnarray}
{d\sigma\over d\Omega}
&=& \left({r_0 m_e \over \hbar}\right)^2 \omega^4 \left|
\epsilon^{(\alpha')}\cdot\epsilon^{(\alpha)}\right.
\nonumber \\
& & \sum_I \left[{|<A\parallel r\parallel I>|^2 \over
\omega_{IA}} 
(1+{\omega\over\omega_{IA}}+{\omega^2\over \omega_{IA}^2}+ \cdots)\right.
\nonumber \\ 
&+& \left.\left.  
{|<I\parallel r\parallel A>|^2 \over\omega_{IA}}
(1-{\omega\over\omega_{IA}} +{\omega^2 \over \omega_{IA}^2} -\cdots)
\right]\right|^2
\nonumber \\
&=& \left({2r_0 m_e \over \hbar}\right)^2 \omega^4 \left|
\epsilon^{(\alpha')}\cdot\epsilon^{(\alpha)}\right|^2
\nonumber \\
& & \left|\sum_I {|<A\parallel r\parallel I>|^2 \over
\omega_{IA}} 
(1+{\omega^2\over\omega_{IA}^2}+{\omega^4\over \omega_{IA}^4}+ \cdots)
\right|^2
\end{eqnarray}
Here, we used a commutation relation 
\begin{equation}
<A|{\bf p}|I>=im_e\omega_{IA}<A|{\bf x}|I>,
\end{equation}
which results from the commutation relation ${\bf p}=m[{\bf x}, H_0]/i\hbar$
with $H_0$ being the Hamiltonian.
The leading term is proportional to $\omega^4$, which is well-known
result for the Rayleigh scattering cross section in the low energy limit.

By performing the angular integration for unpolarized incident radiation
and averaging over the polarization for an outgoing radiation, we have
$\int d\Omega |\epsilon^{(\alpha')}\cdot\epsilon^{(\alpha)}| =
{8\pi\over 3}$. 
We define the dimensionless angular frequencies
\begin{equation}
\tilde\omega_{IA}\equiv {\omega_{IA}\over\omega_l},
\end{equation}
where $\omega_l$ is the angular frequency corresponding to the Lyman limit.
From this we have 
$\tilde\omega_{np\ 1s} =1- n^{-2}$ and
$\tilde\omega_{n'p\ 1s} =1+ (n')^{-2}$
for a bound $np$ state and a continuum $n'p$ state, respectively.  
In a similar way, we define a dimensionless position operator 
\begin{equation}
{\bf \tilde r} = {{\bf r}\over a_0}
\end{equation}
where $a_0=\hbar^2/(m_e e^2)=0.53{\rm\ \AA}$ is the Bohr radius for hydrogen.

With these definitions we may write the total scattering cross section 
\begin{eqnarray}
\sigma(\omega) &=& \sigma_T
\left({\omega\over 3\omega_l}\right)^4
\left|
\sum_I {|<A\parallel {\tilde r}\parallel I>|^2 \over \tilde\omega_{IA}} 
\right. \nonumber \\
& & \left[1+{1\over\tilde\omega_{IA}^2}\left({\omega\over\omega_l}\right)^2
\right.  +
\left.\left. {1\over\tilde\omega_{IA}^4}\left({\omega\over\omega_l}\right)^4
+\cdots \right]
\right|^2.
\end{eqnarray}
Here, $\sigma_T={8\pi r_0^2/3}=0.665\times 10^{-24}{\rm\ cm^2}$ is the Thomson
scattering cross section.
We note that the leading term is proportional to $\omega^4$, which is 
a well-known result for Rayleigh scattering. We are interested in
the higher order terms of $\omega/\omega_l$ for better approximation
redward of Ly$\alpha$.

The matrix elements of the dipole operators 
$<np\parallel \tilde r\parallel 1s>, <n'p\parallel x\parallel 1s>$ are 
easily found in textbooks on quantum mechanics 
(e.g. Berestetski, Lifshitz \& Pitaevskii 1971, Saslow
\& Mills 1969, Bethe \& Salpeter 1967).  The matrix elements are given by
\begin{equation}
<np\parallel \tilde r\parallel 1s> =\left[ {2^{8}n^7 (n-1)^{2n-5}}\over
{3(n+1)^{2n+5}} \right]^{1\over2} 
\end{equation}
for the bound states.  For the continuum states, the corresponding 
values are given by
\begin{equation}
<n'p\parallel \tilde r\parallel  1s>
=\left[ {2^{8}(n')^7 e^{-4n'\tan^{-1}(1/n')}}\over
{3[(n')^2+1]^{5}[1-e^{-2\pi n'}]} \right]^{1\over2} .
\end{equation}
Here, the normalization condition for continuum wavefunctions is
\begin{equation}
\int_0^\infty R_{n_1'p}(r) R_{n_2'p} r^2 dr = \delta(n_1'-n_2'),
\end{equation}
where $\delta(n')$ is the Dirac delta function.

With these matrix elements we may divide the terms involving bound states
and continuum states and write the scattering cross section as a sum
of two power series of $\omega/\omega_l$
\begin{equation}
\frac{\sigma(\omega)}{\sigma_{T}}=\biggr(\frac{\omega}{\omega_{l}}\biggr)^{4} 
\left[
  \sum^{\infty}_{p=0}a_{p} \biggr(\frac{\omega}{\omega_{l}}\biggr)^{2p}
  + \sum^{\infty}_{p=0}b_{p}
  \biggr(\frac{\omega}{\omega_{l}}\biggr)^{2p} \right]^{2}.
\end{equation}
Here, the coefficients $a_p, b_p$ are given by
\begin{eqnarray}
a_{p}&=& \sum^{\infty}_{n=2} \frac{2^{8}n^7 (n-1)^{2n-5}}{3(n+1)^{2n+5}} 
\left(1-\frac{1}{n^2}\right)^{-2p-1} ~, \nonumber \\
b_{p}&=&  \int^{\infty}_{0} dn^{\prime} \frac{2^{8}(n')^7 e^{-4n'\tan^{-1}(1/n')
}}
{3[(n')^2+1]^{5}[1-e^{-2\pi n'}]}
\left(1+\frac{1}{n'^2}\right)^{-2p-1} ~.
\end{eqnarray}

We rewrite Eq.~(12) as a single power series
\begin{eqnarray}
\frac{\sigma(\omega)}{\sigma_{T}} &=&
\left({\omega\over \omega_l}\right)^4
\left[
c_0+c_1\left({\omega\over\omega_l}\right)^2
+c_2\left({\omega\over\omega_l}\right)^4
+\cdots
\right] \nonumber \\
&=&\biggr(\frac{\omega}{\omega_{l}}\biggr)^{4}\sum_{p=0}^{\infty} c_{p} 
\biggr(\frac{\omega}{\omega_{l}}\biggr)^{2p},
\end{eqnarray}
which forms a low energy expansion of the Kramers-Heisenberg formula.
The coefficients $c_{p}$ are obtained from the coefficients 
$a_p$ and $b_p$  by the relation
\begin{equation}
c_{p}=\sum^{p}_{q=0} (a_{q}+b_{q}) (a_{p-q}+b_{p-q}).
\end{equation}
In Table~1, we listed the numerical values of the coefficients $a_p, b_p$ and
$c_p$ up to $p=9$. It is noted that the continuum contribution represented
by $b_p$ is small for high orders, but not negligible for low orders and
must be included for accurate calculations.

\begin{table*}
 \centering
 \begin{minipage}{140mm}
  \caption{Numerical Values of the Coefficients $a_p, b_p$ and $c_p$}
  \begin{tabular}{cccccccc}
  \hline
p & $c_p$ & $a_p$ & $b_p$ & p & $c_p$ & $a_p$ & $b_p$  \\
 \hline
0 & 1.26537 & 0.9157 & 0.2092 & 5 & 81.1018 & 13.5826 & 0.0571 \\ 
1 & 3.73766 & 1.52456 & 0.1368 & 6 & 161.896 & 23.9 & 0.04983 \\ 
2 & 8.8127 & 2.58886 & 0.1015 & 7 & 319.001 & 42.19 & 0.0442 \\ 
3 & 19.1515 & 4.4587 & 0.08061 & 8 & 622.229 & 74.641 & 0.03971 \\ 
4 & 39.919 & 7.75546 & 0.06685 & 9 & 1203.82 & 132.251 & 0.03605 \\
\hline
\end{tabular}
\end{minipage}
\end{table*}

\section{Result}
 
\subsection{Low Energy Expansion}

In Fig.~1, we show our result for the low energy expansion 
in Eq.~(14) up to $p=9$ by a long dashed line. In the figure, we also
show the result from the fully quantum mechanical
computation, which was numerically obtained by performing the summation and
integration appearing in the Kramers-Heisenberg formula. 
The exact cross section of Rayleigh scattering by atomic
hydrogen was presented by Gavrila (1967), who computed the Green
function in momentum space and provided the cross section in
a tabular form.  The crosses in the figure
mark the result from the work of Gavrila (1967). It is noted that 
the cross marks are found on the solid line indicating excellent agreement.

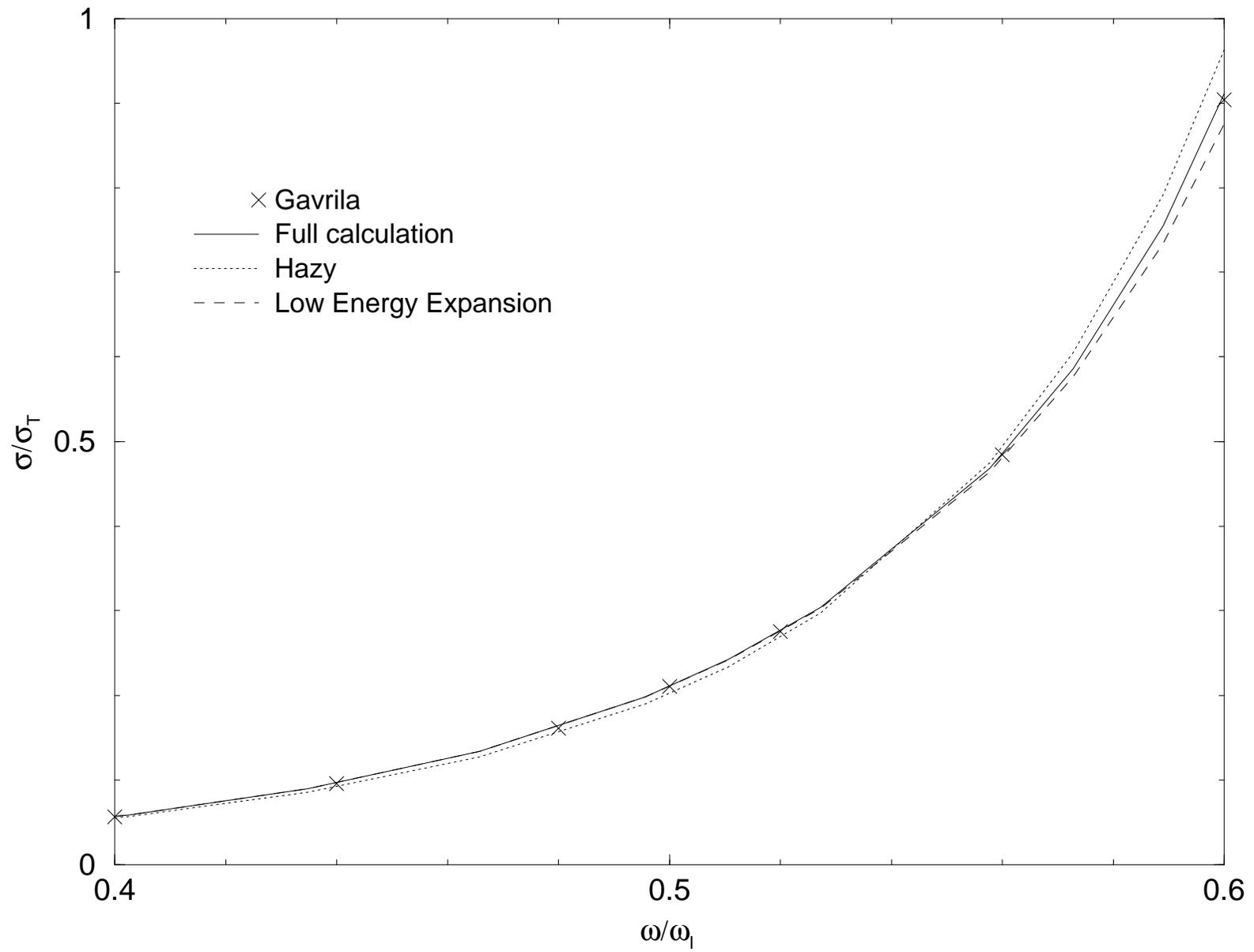
\begin{figure}
 \vspace{302pt}
 \caption{
The cross section of Rayleigh scattering for incident radiation with
angular frequency $0.4\omega_l<\omega<0.6\omega_l$, where $\omega_l$
is the angular frequency corresponding to the Lyman limit.
The result obtained from
the full Kramers-Heisenberg formula is shown by the solid line, 
and the dashed line represents our low energy expansion given in Eq.~(14)
up to $p=9$ (see the text).  The crosses mark the values given in a
tabluar form by Gavrila (1969), which excellently fall on the solid curve.
The polynomial fit (Eq.~16) by Ferland (2001) to the work of Gavrila (1969) 
is depicted by the dotted line.
The dashed line is always located under the solid line, and such a behaviour
is  absent in the case of the polynomial fit given by Ferland, for
which it crosses the solid line at $\omega/\omega_l \sim 0.54$.  }
\end{figure}

Because the Ly$\alpha$ resonance is located at $\omega/\omega_l=0.75$, 
the covergence is rather slow for $\omega/\omega_l \simeq 0.6$,
which requires a large number of terms in the expansion. 
This behaviour is natural because the scattering cross section
changes too steeply near resonance for any polynomial approximation
to be suitable.
On the other hand, in the region with $\omega/\omega_l <0.5$ 
the agreement is almost perfect and the two results are not
distinguished in the figure.

For comparison purpose, we make a plot for the polynomial fit
to the result of Gavrila (1967) obtained by Ferland (2001). 
He incorporated it in his photoionization code `Cloudy,' 
in which the formula is quoted to be useful for radiation with
$\lambda >1410 {\rm\ \AA}$, corresponding to $\omega/\omega_l = 0.647$.  
He used the Lorentzian function instead for 
$\lambda <1410 {\rm\ \AA}$, where the resonance of Ly$\alpha$ dominates.
The polynomial fit to the scattering cross section obtained 
by Ferland (2001) is
\begin{eqnarray}
\sigma_{Cl}(\omega) &=& \Bigg[8.41 \times 10^{-25}
\left({\omega\over\omega_l}\right)^4 +3.37\times 10^{-24}
\left({\omega\over\omega_l}\right)^6 \nonumber \\
&+&4.71\times 10^{-22}
\left({\omega\over\omega_l}\right)^{14} \Bigg]{\rm\ cm^{2}},
\end{eqnarray}
which shows the same $\omega^4$ dependence in the leading term.  
In Fig.~1, we also show the values obtained from this polynomial fit
by a dotted line. 

Because the coefficients $a_p, b_p$ and $c_p$ are all positive, any finite
expansion of Eq.~(14) gives smaller values than the true ones
obtained from the full Kramers-Heisenberg formula. Therefore, the long dashed
line is always located under the solid line. However, such a behaviour
cannot be seen in the case of the polynomial fit given by Eq.~(16), for
which it crosses the solid line at $\omega/\omega_l \sim 0.54$.

\subsection{Near-Resonance Behavior}

Lee (2003) discussed the deviation of the scattering cross section
from the Lorentzian near Ly$\alpha$ resonance. The deviation is attributed
to contributions from infinitely many other states than $2p$. 
According to him, near resonance, the Kramers-Heisenberg formula can be 
expanded in terms of
$\Delta\omega = \omega -\omega_{21}$, where
$\omega_{21} = 0.75\omega_l$ is the angular frequency corresponding
to the Ly$\alpha$ transition. 
Explicitly, the expansion can be written as
\begin{eqnarray}
\sigma(\omega) &=& \sigma_T \left( {\omega_{21}\over\Delta\omega} \right)^2
\left|A_0+A_1\left({\Delta\omega\over\omega_{21}}\right) \right. \nonumber \\ 
&+&\left. A_2\left({\Delta\omega\over\omega_{21}}\right)^2 
  +\cdots \right|^2.  
\end{eqnarray}
The coefficients up to $A_5$ are given by
\begin{eqnarray}
A_0 &=& f_{12}/2 = 0.2081 \nonumber \\
A_1/A_0 & = & -0.8961  \nonumber \\
A_2/A_0 & = & -1.222\times 10^1  \nonumber \\
A_3/A_0 & = & -5.252\times 10^1   \nonumber \\
A_4/A_0 & = & -2.438\times 10^2   \nonumber \\
A_5/A_0 & = & -1.210\times 10^3,   
\end{eqnarray}
where $f_{12}=0.4162$ is the oscillator strength for the Ly$\alpha$
transition.

With these coefficients, the Rayleigh scattering cross section near
Ly$\alpha$ can be written explicitly up to the fifth order
\begin{eqnarray}
{\sigma_{L}(\tilde{\omega})\over \sigma_T}&=& {0.0433056 \over 
\tilde{\omega}^{2}} (1-1.792
\tilde{\omega} - 23.637 \tilde{\omega}^{2} \nonumber \\ 
&-& 83.1393 \tilde{\omega}^{3} - 244.1453 \tilde{\omega}^{4} -
699.473\tilde{\omega}^{5} )
\end{eqnarray}
where we define $\tilde{\omega}\equiv(\omega-0.75)/0.75=
\Delta\omega/\omega_{21}$.

In Fig.~2 we make a plot of this expansion up to the fifth order represented
by a dotted line. We also compare this result with
the results from the full Kramers-Heisenberg formula and approximations
with lower order corrections. In this figure, the cross marks from
the result of Gavrila (1969) fall on the solid line representing
the full Kramers-Heisenberg formula as seen in Fig.~1.
The agreement with the fifth order correction to the Lorentzian
is excellent.

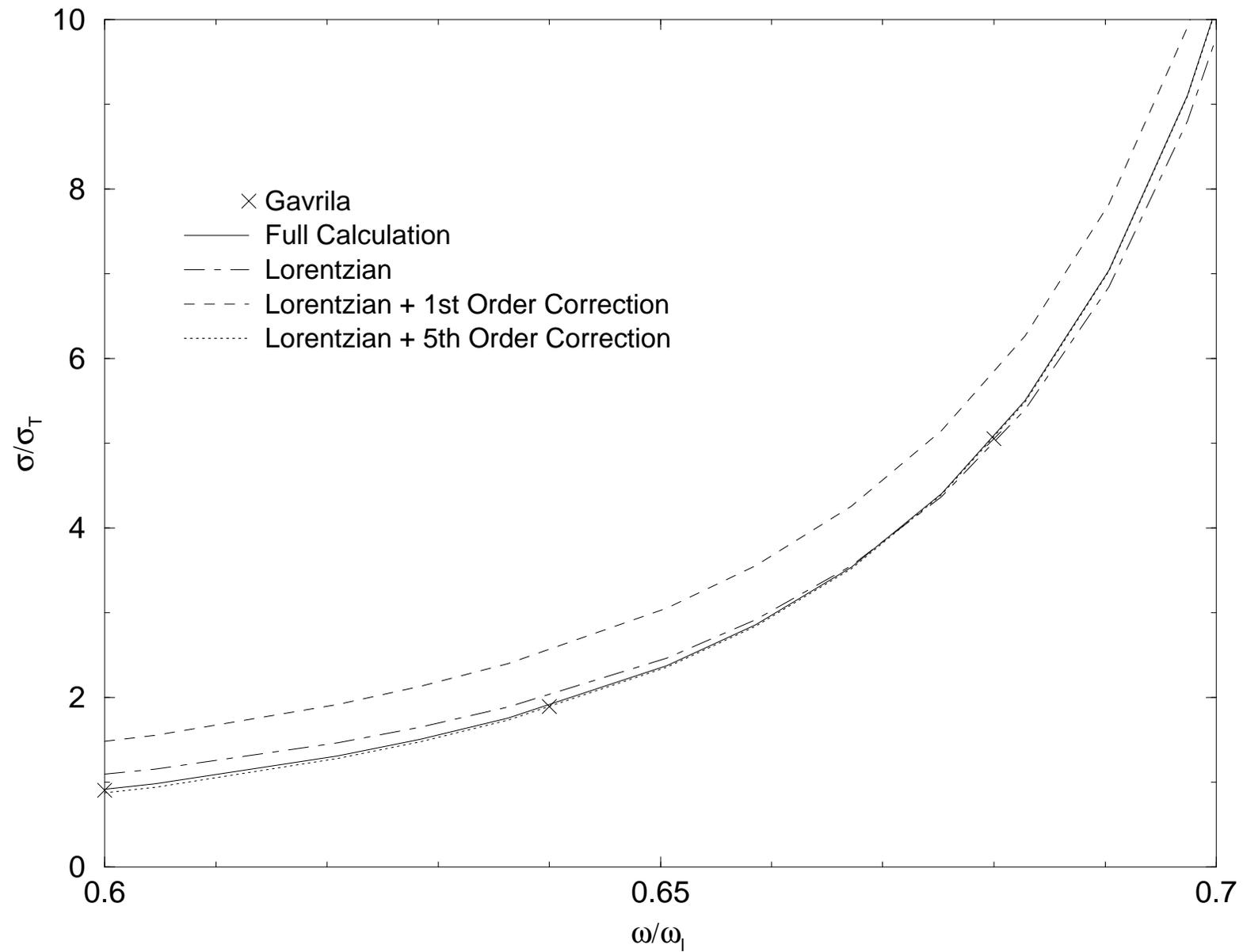
\begin{figure}
 \vspace{302pt}
 \caption{
The cross section of Rayleigh scattering for incident radiation with
angular frequency $0.6\omega_l<\omega<0.7\omega_l$. Similarly with Fig.~1,
the solid line represents the full Kramers-Heisenberg formula and
the crosses are the values appearing in Gavrila (1969). The dotted line
represents the Lorentzian with the fifth order correction (Eq.~19), 
which gives excellent agreement with the full Kramers-Heisenberg formula.
In the code `Cloudy' developed by Ferland (2001), a simple Lorentzian
is used, which also is a good approximation, represented by the dot-dashed
line. The dashed line represents the first order correction, which
is poorer than the simple Lorentzian in this particula region. However,
Lee (2003) showed that the first order correction is much better 
in the region nearer the Ly$\alpha$ resonance.}
\end{figure}

It is very interesting to note that the approximation of the first 
order correction to the Lorentzian depicted by a dashed line is poorer 
than just the 
Lorentzian without any correction terms in the region $\omega/\omega_l
\sim 0.7$ represented by a dot-dash line.  
However, the first order approximation is excellent and better than
the Lorentzian very near the resonance, i.e., 
$\frac{\omega}{\omega_{l}} > 0.70$, which is not plotted here 
(see Lee 2003). This behaviour is explained 
by the alternating nature of the series given in Eq.~(19).

As is done in the photoionization code `Cloudy' by Ferland (2001), 
the Rayleigh scattering cross section
is approximated by the Lorentzian near resonance $\omega/\omega_l >0.647$
and Eq.~(16) for $\omega/\omega_l<0.647$, which gives accurate results within
errors not exceeding 5 \%. In a similar way, the combination of Eq.~(14)
up to $p=9$ for $\omega/\omega_l<0.6$ and Eq.~(19) for $\omega/\omega_l>0.6$
is accurate within errors less than 4 \%. In particular, it should be
emphasized that significant errors are found only near the boundary 
region dividing the two expansions. 

\section{Discussion and Observational Ramifications}

In this paper, we obtained an expansion in terms of 
$\omega/\omega_l$ of the Rayleigh scattering cross section by atomic hydrogen,
which is applied in the low energy regime with $\omega/\omega_l<0.6$.
By combining this with another expansion of the Kramers-Heisenberg formula 
around the Ly$\alpha$ resonance in
terms of $\Delta\omega=(\omega-\omega_{Ly\alpha})/\omega_{Ly\alpha}$,
we may have a wieldy and useful approximate formula 
for the Rayleigh scattering process redward of Ly$\alpha$ 
by atomic hydrogen, which can be made arbitrarily accurate
by inclusion of higher order terms directly calculated from the
Kramers-Heisenberg formula.

Rayleigh scattering by atomic hydrogen is important only in the presence
of a scattering region with a very high neutral hydrogen column density
$N_{HI}$.  Such high column density media may be found in an extended 
atmosphere of a giant star where the mass loss process is already very 
important.  Isliker et al. (1989) considered the effect 
of Rayleigh scattering in binary systems containing a giant star.
In these systems, for a given inclination and density distribution, 
the scattering optical depth
is dependent on the wavelength, and therefore light curves
differ according to the observed wavelength. This information may be
quite important to investigate the mass loss process from a giant star.

In their analysis, Isliker et al. (1989) presented the Rayleigh 
scattering cross section given by
\begin{equation}
\sigma(\omega) = \sigma_T \left[\sum_{k=2}^\infty {f_{1k}\over 
\left({\omega_{1k}\over \omega}\right)^2-1}\right]^2,
\end{equation}
where $f_{1k}$ is the oscillator strength between $1s$ and $kp$ states.
In their work, they neglected the contribution from the continuum states.
However, as is noted in the previous section, the contribution from
the continuum states to the low energy regime is not negligible, and
hence caution should be exercised.

In more than half of the symbiotic stars, Raman scattered O~VI 6827, 7088
features are seen, which are formed via Raman scattering of O~VI 1032, 1038 
resonance doublet by atomic hydrogen. Being slightly less energetic than 
Ly$\beta$, O~VI 1032, 1038 doublet may excite a hydrogen atom that can
subsequently de-excite to excited $2s$ state re-emitting an optical photon
redward of H$\alpha$. This process was first identified by Schmid (1989),
where the scattering cross section is of similar order to that for Rayleigh
scattering (e.g. Lee \& Lee 1997, Nussbaumer,  Schmid, \& Vogel 1989).

A very high column density media may be found in an early universe
when the reionization of intergalactic medium initiated by the first objects 
was not completed. In this partially ionized universe, a significant
extinction around Ly$\alpha$ is expected, which will result in a big 
absorption trough known as the Gunn-Peterson effect (Gunn \& Peterson
1965, Scheuer 1965). Thus far several quasars with redshift $z>6.2$ have
been idenfied with Gunn-Peterson troughs (Becker et al. 2001, 
Fan et al. 2003).  It appears that the H~I 
column density that is responsible for these Gunn-Peterson troughs
are not sufficiently high for applications of our current work, but
may be high enough to see the deviation from the Lorentzian approximation
of the scattering cross section (see Lee 2003, Miralda-Escud\'e 1998).
Deeper IR search for higher redshifted objects may exhibit extremely
high neutral hydrogen column density, where an accurate calculation
of the cross section is required.

\section*{Acknowledgments}
This work is a result of research activities of the Astrophysical
Research Center for the Structure and Evolution of the Cosmos (ARCSEC)
funded by the Korea Science and Engineering Foundation.

\label{lastpage}


\begin{thebibliography}{99}

\bibitem[Becker et al. 2001]{bec01} Becker, R. H. et al. 2001, AJ, 122, 2850
\bibitem[Berestetskii, Lifshitz, \& Pitaevskii, 1971]{ber71}
Berestetskii,V.B., Lifshitz, E.M., \& Pitaevskii, L.P., 1971,
Relativistic Quantum Theory, Pergamon Press
\bibitem[Bethe \& Salpeter 1967]{bet67} Bethe, H. A. \& Salpeter, E. E.  1967,
Quantum Mechanics of One and Two Electron Atoms, Academic Press Inc., New York
\bibitem[Fan et al. 2003]{fan03} Fan, X. et al. 2003, AJ, 125, 1649
\bibitem[Ferland 2001]{fer01} Ferland, G., 2001, {\it Hazy, a brief
introduction to Cloudy 94.00}
\bibitem[Gavrila 1967]{gav67} Gavrila, M., 1967, Physical Review, 163, 147
\bibitem[Gunn \& Peterson 1965]{gun65} Gunn, J. E., Peterson, B. A., 1965,
ApJ, 142, 1633
\bibitem[Isliker, Nussbaumer \& Vogel 1989]{isl89}Isliker, H.,
Nussbaumer, H., \& Vogel, M., 1989, A\&A,  219, 271
\bibitem[Lee 2003]{lee03} Lee, H. -W. 2003,
ApJ, 594, 627
\bibitem[Lee \& Ahn 1998]{lee98} Lee, H. -W., \& Ahn, S. -H. 1998,
ApJ, 504, L61
\bibitem[Lee \& Lee 1997]{lee97} Lee, H. -W., \& Lee, K. W. 1997,
MNRAS, 287, 211
\bibitem[Merzbacher 1961]{mer61} Merzbacher, E. 1970, Quantum Mechanics,
Wiley, New York
\bibitem[Miralda-Escud\'e 1998]{mir98} Miralda-Escud\'e, J., 1998, ApJ,
501, 15
\bibitem[Nussbaumer, Schmid, \& Vogel 1989]{nus89}  Nussbaumer, H.,
Schmid, H. M.\& Vogel, M.,1989,A\&A, 221, L27
\bibitem[Sakurai 1967]{sak67} Sakurai, J. J., 1967, Advanced Quantum
Mechanics, Addison-Wesley Publishing Company, Reading, Massachusetts
\bibitem[Saslow \& Mills 1969]{sas69} Saslow, W. M., Mills, D. L. 1969,
Physical Review, 187, 1025
\bibitem[Scheuer 1965]{sch65} Scheuer, P. A. G. 1965, Nature, 207, 963
\bibitem[Schmid 1989]{sch89} Schmid, H. M. 1989,A\&A,  211, L31
\end{thebibliography}
\end{document}